\documentclass[runningheads]{llncs}
\usepackage{amssymb}\usepackage{tikz}
\usetikzlibrary{arrows,automata,shapes,positioning}
\usetikzlibrary{positioning,decorations.markings}
\usepackage{fullpage}
\newcommand{\myparagraph}[1]{\emph{#1}}
\usepackage{thmtools,thm-restate,todonotes}
\newcommand{\set}[1]{\{#1\}}

\newcommand{\aut}{\mathcal{A}}
\newcommand{\autB}{\mathcal{B}}

\newcommand{\Z}{\mathbb{Z}}

\newcommand{\buchi}{B\"uchi{}}

\newcommand{\PTIME}{\textsc{PTime}{}}

\newcommand{\NP}{\textsc{NP}{}}
\newcommand{\coNP}{\textsc{coNP}{}}

\newcommand{\cost}{\mathbf{c}}

\newcommand{\tuple}[1]{\langle #1 \rangle}

\newcommand{\lang}{\mathcal{L}}

\tikzset{
    position/.style args={#1:#2 from #3}{
        at=(#3.#1), anchor=#1+180, shift=(#1:#2)
    }
}

\usepackage{etoolbox,refcount}
\usepackage{multicol}

\newcounter{countitems}
\newcounter{nextitemizecount}
\newcommand{\setupcountitems}{\stepcounter{nextitemizecount}\setcounter{countitems}{0}\preto\item{\stepcounter{countitems}}}
\makeatletter
\newcommand{\computecountitems}{\edef\@currentlabel{\number\c@countitems}\label{countitems@\number\numexpr\value{nextitemizecount}-1\relax}}
\newcommand{\nextitemizecount}{\getrefnumber{countitems@\number\c@nextitemizecount}}
\newcommand{\previtemizecount}{\getrefnumber{countitems@\number\numexpr\value{nextitemizecount}-1\relax}}
\makeatother

\newcommand{\podwa}{{\Lambda}} \title{Deterministic Weighted Automata under Partial Observability}
\author{Jakub Michaliszyn\inst{1}\orcidID{0000-0002-5053-0347} \and
Jan Otop\inst{1}\orcidID{0000-0002-8804-8011}}
\authorrunning{J. Michaliszyn and J. Otop}
\institute{University of Wroc\l{}aw\\
\email{\{jmi,jotop\}@cs.uni.wroc.pl}}

\begin{document}
\maketitle
\begin{abstract}
Weighted automata is a basic tool for specification in quantitative verification, which allows to express quantitative features of analysed systems such as resource consumption.
Quantitative specification can be assisted by automata learning as there are classic results on Angluin-style learning of weighted automata.
The existing work assumes perfect information about the values returned by the target weighted automaton.
In assisted synthesis of a quantitative specification, knowledge of the exact values is a strong assumption and may be infeasible. 
In our work, we address this issue by introducing a new framework of \emph{partially-observable} deterministic weighted automata,
in which weighted automata return intervals containing the computed values of words instead of the exact values. 
We study the basic properties of this framework with the particular focus on the challenges of active learning {partially-observable} deterministic weighted automata.
\end{abstract}

\section{Introduction}
Finite automata is a fundamental computational model with a wide range of applications spanning from
computational complexity, through AI~\cite{millington2019ai} to formal methods~\cite{clarke2018handbook}.
In some applications, however, the qualitative answers returned by finite automata, i.e., each word is \emph{accepted} or \emph{rejected}, 
are insufficient.
For instance, in formal verification, one can check the existence of execution trances violating a given specification, 
but violating traces come from a model rather than the actual system and their severity may differ from critical, 
which are likely to occur in the actual system to one, which are unlikely to be reproduced. 
Similarly, while checking whether a system has no deadlocks, one can ask whether every request is eventually fulfilled, 
which lacks performance guarantees involving a bound on the timeframe for fulfilment.

To address these issues, there has been proposed quantitative verification, in which the specification refers to quantitative features of the system. 
Quantitative verification is based on weighted automata, which return numeric values for words rather than accept/reject words. 
Weighted automata and their extensions have been extensively studied~\cite{WABook,QL,NWA}. 
These models can express the severity of errors~\cite{modelMeasuring} and various performance metrics such as average response time~\cite{NWA}.
The expressive power of such models entails hardness of specification. 

Specifying quantitative properties may be difficult because in addition to describing events (such as a system failure) one has to come up with the associated values. 
This is especially difficult for properties  of an approximate nature such as the aforementioned severity of a failure. 
Furthermore, the precise values are often not that important as we would be typically interested whether the number is within some acceptable interval, 
e.g., does not exceed our resources. 
For instance, the exact value of average response time depends on the computing environment, e.g., its cache size, which is typically not modeled precisely. 
For the same reason, assigning reasonable values of the average response time to traces is considerably more difficult than specifying a deadlock. 

In this paper, we address the issue of construction of quantitative specifications. 
To ease the specification process, we propose a new framework, in which automata do not reveal the exact values.
We study this framework from the specification-synthesis perspective
, i.e., we ask whether it is possible to 
semi-automatically produce quantitative specifications using automata-learning approach. 
The conditions may be more involved; for example, we may want to express properties stating that the values 0-10 are good, 11-20 are satisfactory, and anything over 20 is bad.

\subsection{Our framework}
We introduce partially-observable deterministic weighted automata (PODWA). 
These automata behave as regular deterministic weighted automata over $\Z$, but return an interval (from a given finite set of possible intervals) that contains 
the computed value rather than the value itself. The choice of intervals as partial observations is natural. 
Other choices are possible, but can increase the complexity -- even making the membership problem undecidable.

Our motivation comes from the specification-synthesis via automata learning. 
The idea is that we would like to be able to synthesize quantitative properties without necessarily providing exact values. 
For that reason, we focus on problems related to active automata learning. 
First, we study the equivalence problem. It is fundamental in automata learning as one needs to answer whether the learned automaton is admissible. 
Second, learning algorithms typically construct the structure of an automaton with no weights~\cite{learningWA}, which leads to the weight synthesis question: 
given a PODWA $\podwa_1$ and an automaton structure $\aut_2$ (a deterministic finite automaton) without weights, 
is there a weight assignment for $\aut_2$, which makes it equivalent (w.r.t. partial observations) to $\podwa_1$?
Specifically, assuming that such a weight assignment does exist, is there one such that weights vales are of polynomial order w.r.t. weights from $\podwa_1$?
Finally, active automata learning algorithms construct minimal automata~\cite{Lstar,learningWA,learningTA}. Thus, to assess feasibility of learning weighted automata in our framework, we
study the minimization problem for PODWA.

\subsection{Results}
The main contribution of the paper is identifying obstacles in developing a polynomial-time active learning algorithm for the new model. 
We start with the basic properties of the model. 
We show that the class of PODWA can express more than regular languages and is closed under the complement, but not under the union or the intersection. Then, we show that:
\begin{itemize}
\item~the equivalence problem for PODWA is \coNP-complete in general, and it can be solved in polynomial time if weights are given in unary,
\item~there is a PODWA $\podwa$ with weights $-1,0,1$, such that all equivalent minimal-state automata are isomorphic and have exponential weights, and
\item~the minimization via state-merging for PODWA with unary weights is \NP-complete.
\end{itemize}

These results highlight challenges in learning weighted automata under partial observation. 
In order to obtain polynomial-time algorithm for active learning of PODWA, we need to focus on automata with unary weights. 
However, equivalence up to partial observation is too permissive to have an active learning algorithm. 
One needs a more rigid equivalence notion, which would make minimization decidable in polynomial time, and prevent exponential blow-up of weights 
in the minimization process. 

\subsection{Related work}  
Typically, the partial observation term applies to equivalence on the set of control states, which has been used to model decisions under imperfect information in 
Markov decision processes (partially observable Markov decision process~\cite{DBLP:journals/mor/PapadimitriouT87}), 
graph games (games with imperfect information~\cite{imperfectInformationGames}), or
multi-agent system (multi-player games with imperfect information~\cite{blackburn2006handbook,guelev2012epistemic}). In contrast, in this work, the state space is intact, and partial observability refers to 
the returned value. This is related to games with interval objectives~\cite{DBLP:conf/fsttcs/HunterR14}, in which one of the players objective is to 
make the numeric outcome of the game fall into a set being a finite union of intervals.

This work is motivated by active automata-learning algorithms, which have been developed for deterministic finite automata~\cite{Lstar}, 
deterministic weighted word automata~\cite{learningWA} and deterministic weighted tree automata~\cite{learningTA} and other types of automata. 
Similar algorithms have recently been developed for infinite-word  automata: deterministic \buchi{}  automata (DBA)~\cite{DBLP:journals/ai/MichaliszynO22} and 
deterministic limit-average automata~\cite{DBLP:conf/ijcai/MichaliszynO21}. 
These algorithms work in polynomial time even though minimization, closely related to active learning, is \NP-complete for DBA.
It was made possible thanks to in-depth difficulty assessment of problems related to active learning, which indicated how to extend the learning framework to make
polynomial-time learning algorithms possible~\cite{DBLP:journals/ai/MichaliszynO22}. 
We conduct such an assessment in this work to pave the way for the development of active learning algorithms.

\section{Preliminaries}
A \emph{word} $w$ is a finite sequence 
of letters from a finite alphabet $\Sigma$.
By $\Sigma^*$ we denote the set of all words over $\Sigma$.
By $w[i]$ we denote the $i$th letter of a word $w$, and $w[i,j]$ stands for the subword $w[i] w[i+1] \ldots w[j]$ of $w$. 
The empty word is denoted by $\epsilon$.

\myparagraph{Automata and runs.} 
A \emph{deterministic weighted automaton} (DWA) is a tuple $\tuple{\Sigma, Q, q_0, \delta, \cost}$
consisting of
\begin{enumerate}
\item	~an alphabet $\Sigma$, 
\item	~a finite set of states $Q$, 
\item	~an initial state $q_0 \in Q$,  
\item	~a transition function 	$\delta \colon Q \times \Sigma \to Q$, and 
\item	~a weight function $\cost \colon Q \times \Sigma \to \Z$.
\end{enumerate}
The size of a DWA $\aut$, denoted by $|\aut|$, is its number of states plus the sum of the lengths of all the weights given in binary.

We extend $\delta$ to $\hat{\delta} \colon Q \times \Sigma^* \to Q$  inductively: for each $q$, we set $\hat{\delta}(q, \epsilon) = q$, and 
for all $w\in \Sigma^*, a \in \Sigma$, we set $\hat{\delta}(q, wa)=\delta(\hat{\delta}(q, w), a)$.
The run $\pi$  of a DWA $\aut$ on a word $w$ is the sequence of states $q_0\hat{\delta}(q_0, w[1]) \hat{\delta}(q_0, w[1,2]) \dots$.
We do not consider any acceptance condition here.

The semantics of a DWA $\aut$ is a function $\lang(\aut)$ from non-empty words $\Sigma^* \setminus \{\epsilon\}$ into integers. 
For a non-empty word $w$ of length $k$, we define $\lang(\aut)(w)$ as the sum of weights of transitions along the run of $\aut$ on $w$:
\[
\lang(\aut)(w) = \cost(q_0, w[1]) + \cost(\hat{\delta}(q_0, w[1]), w[2]) + \dots + \cost(\hat{\delta}(q_0, w[1, k-1]), w[k]).
\]

\begin{remark}{The tropical seminring}
The weighted automata model considered in this paper is an instance of a more general framrework of weighted automata over semirings~\cite{WABook}, where the semiring is 
the tropical semiring restricted to integers.
\end{remark}

\section{Our framework}

A \emph{Partially-Observable DWA}, PODWA, is a pair $\podwa = (\aut, S)$ consisting of 
a DWA $\aut$ and a set of a finite number of pairwise-disjoint intervals $S$ covering $\Z$ called \emph{observations}. 
We assume that intervals are enumerated by $\set{0, \dots, s}$ according to the order on $\Z$.
The \emph{language} of a PODWA $\podwa$, denoted as $\lang(\podwa)$, is a function from $\Sigma^* \setminus \set{\epsilon}$ to $\set{0, \dots, s}$ such that $\lang(\podwa)(w)$ is the number of the interval containing $\lang(\aut)(w)$.

A \emph{binary PODWA} is a special case of PODWA having only two intervals: 
$(-\infty, 0]$ and $(0, +\infty)$. We consider words ending in the interval $(0, +\infty)$ as \emph{accepted}. 
Then, the function $\lang(\podwa)$ is essentially a characteristic function of a set that can be seen as a classic language.

\begin{example}\label{e:morecthana}
Consider a single-state automaton $\aut$ over $\Sigma = \set{a,b,c}$. The weights of the transitions over $a, b, c$ are, respectively, $-1, 0, 1$.
Consider the set of intervals $S = \set{(-\infty, 0], (0, +\infty)}$ and the binary PODWA $\podwa = (\aut, S)$. 
Then, $\lang(\podwa)(w) = 1$ if $w$ contains more occurrences of $c$ than $a$, and $0$ otherwise.
\end{example}

Binary PODWA can define all regular languages (without the empty word) and some non-regular languages (see Example~\ref{e:morecthana}).
All PODWA-recognizable languages are context-free and can be emulated by a deterministic one-counter automaton. 
On the other hand, deterministic one-counter automata define languages that cannot be expressed by binary PODWA, as the former 
rely on the counter value at every transition while the latter are agnostic of the counter value.
For instance, a pumping argument shows that the language of words 
that have the same number of (occurrences of) $a$ and $b$ between every pair of $c$ 
cannot be expressed by a binary PODWA (or any other PODWA with a reasonable language definition). 

Binary PODWA can be easily complemented -- it suffices to multiply all the weights by $-1$ and adjust the initial state (for words with value $0$). 
We show that the class of languages recognizable by binary PODWA is not closed under union nor intersection. 
We will prove the former; for the latter observe that closure under intersection implies closure under union as the union operation can be expressed by 
the intersection and complement operations.

Let $\lang_\cup$ be the language of words $w$ that the number of occurrences of $c$ is greater than the number of occurrences of $b$ or is greater than the number of occurrences of $a$.
Observe that $\lang_\cup$ is the union of two PODWA-recognizable languages $\lang_a$, $\lang_b$, they can be defined as in Example~\ref{e:morecthana}.
A simple pumping argument shows that $\lang_\cup$ is not PODWA-recognizable.

\begin{restatable}{lemma}{Union}
\label{l:union}
$\lang_\cup$ is not PODWA-recognizable.
\end{restatable}

\begin{proof}
Assume a PODWA $\podwa = (\aut, \set{(-\infty, 0], (0, +\infty)})$ with less than $N$ states that recognizes $\lang_\cup$. 

Consider the word $w = a^{N} b^{N} c^{N+1}$. 
Clearly, $w \in \lang_\cup$ because there are more occurrences of $c$ than $a$.

Since $\podwa$ has less that $N$ states, there is $k \geq 0$ and $l > 0$ with $k+l \leq N$ such that the states $\hat{\delta}(q_0, a^k)$ and $\hat{\delta}(q_0, a^{k+l})$ are the same.

Since the automaton is deterministic, for any $j$ the states $\hat{\delta}(q_0, a^k)$ and 
$\hat{\delta}(q_0, a^{k+jl})$ are the same.
Notice that since the automaton is deterministic, this implies that for any $j$ we have $\hat{\delta}(q_0,  a^{N}) = \hat{\delta}(q_0,  a^{N+j\cdot l})$.

Let $w_i = a^{k + i \cdot l}$.
We argue that $\aut(w_1)-\aut(w_0) \geq 0$. 
Notice that for any $j$ we have $\aut(w_{j+1})-\aut(w_j) = \aut(w_1)-\aut(w_0)$.
If this number was negative, for a sufficiently large $j$ we would have 
\[\aut(a^{N+jl} b^{N} c^{N+1}) \leq 0\]
which contradicts the fact that this words belongs to $\lang_\cup$.

Similarly, there is $k' \geq 0$ and $l' > 0$ with $k'+l' \leq N$ such that the states $\hat{\delta}(q_0, a^{N} b^{k'})$ and $\hat{\delta}(q_0, a^{N} b^{k'+l'})$ are the same.

Let $w'_i = a^N b^{k'+i\cdot l'}$. 
As before, we can show that \(\aut(w'_1) - \aut(w'_0) \geq 0\).

Now consider $w_F = a^{N+l} b^{N+l'} c^{N+1}$. The above reasoning shows that $\aut(w_F) \geq \aut(w)$. 
However, since $\podwa$ recognizes $\lang_\cup$, we have $\aut(w_F) \leq 0$ and $\aut(w)>0$, which is a contradiction.
 \qed
\end{proof}

\subsection{Sample fitting}
We briefly discuss the following counterpart of the sample fitting problem, which is related to passive learning: 
given a set of pairs consisting of a word and an interval, called \emph{the sample}, and a number $n$, 
is there a PODWA with $n$ states that is consistent with the sample? 
The sample fitting problem is \NP{}-complete for PODWA; it is \NP-complete even for DFA. 
However, we discuss it here as the hardness proof is simple and robust. 

For membership in \NP, observe that if $n$ is larger than the number of letters in the sample 
(and the sample does not contain a direct contradiction, i.e., a word with different intervals), 
then such a PODWA always exists (and can be a tree). Otherwise, we can nondeterministically pick a PODWA and check it in polynomial time. 

For hardness, consider an instance $\varphi$ of $3$-SAT with variables $p_1, \dots, p_m$. 
Consider $n=1$, $\Sigma=\set{q} \cup \set{p_i, \overline{p_1} \mid i \leq m}$,
and $S = \set{(-\infty, 0), [0, 1], [2, +\infty)}$. 
The sample consists of:
\begin{itemize}
\item $(q, [0, 1])$, $(qq, [2, +\infty))$
\item $(p_i, [0, 1])$,
$(\overline{p_i}, [0, 1])$,
$(p_i \overline{p_i}, [0, 1])$
$(p_i \overline{p_i} q, [2, +\infty))$ for each $i$
\item $(x y z q, [2, +\infty))$ for each clause $x \lor y \lor z$ of $\varphi$ (we identify $\neg p_i$ with $\overline{p_i}$).
\end{itemize}

If there is a single-state automaton  consistent with this sample, then each letter has a value corresponding 
to the only transition over this letter. The value of each letter is an integer.
The first condition guarantees that the value of $q$ is $1$.
The second guarantees that exactly one letter among $p_i$, $\overline{p_i}$ has value $1$ and the other has the value $0$ 
(we rely on the fact that the weights are over integers).
Thus, the values define a valuation of variables $p_1, \dots, p_m$ from $\varphi$. 
The last condition guarantees that this valuation satisfies every clause of $\varphi$, and thus it satisfies $\varphi$.

\section{Towards active learning PODWA}
The sample fitting problem is intractable for one-state automata, which is a strong negative result for passive learning. 
In this section, we now focus on active learning of automata. 
The classic $L^*$-algorithm for active learning of DFA asks membership and equivalence queries. 
While in the PODWA framework, answering a membership query  amounts to evaluating the DWA over the input word and returning the interval containing the value, 
answering equivalence queries is more involved.

\subsection{Equivalence}
PODWA $\podwa_1$, $\podwa_2$ are \emph{equivalent} if $\lang(\podwa_1) =  \lang(\podwa_2)$. 
The sets of intervals may be different and hence PODWA equivalence is invariant to linear operations, which are consistently applied to all weights and intervals. 
The \emph{equivalence problem} asks whether two given PODWAs are equivalent. 
We show its \coNP-hardness via reduction from (the complement of) the subset sum problem~\cite{HU79}. 
Let $a_1, \ldots, a_k$ be a list of integers and $T$ be the target value represented in binary. W.l.o.g. we assume that $a_1, \ldots, a_k$ are even.
We construct two binary PODWA $\podwa_1 = (\aut_1, S), \podwa_2=(\aut_2, S)$ (where $S = \set{(-\infty, 0], (0, +\infty)}$) such that $\aut_1$ computes the possible values of sums of subsets of $\set{a_1, \ldots, a_k}$ 
minus $T$, and $\aut_2$ returns the value in $\aut_1$ plus $1$, i.e., $\lang(\aut_2)(w) = \lang(\aut_1)(w) +1$.
Observe that $\podwa_1$ and $\podwa_2$ are not equivalent if and only if $\aut_1$ returns $0$ for some word.
For such a word $\aut_2$ returns $1$, which is in a different interval than $0$. 
Thus, the PODWAs are not equivalent if and only if the subset sum problem has a solution.

\begin{restatable}{lemma}{Eqhard}
The equivalence problem for (binary) PODWA is \coNP-hard.
\label{l:eq-hard}
\end{restatable}
\begin{proof}
We discuss the construction of DWA $\aut_1, \aut_2$ such that PODWA
$(\aut_1, S)$ and $(\aut_2,S)$ are equivalent if and only if there is no subsequence of $a_1, \ldots, a_k$, which sums up to $T$.

Without loss of generality, we assume that all values $a_1, \ldots, a_k$ and $T$ are even.
The automaton $\aut_1$ works over the alphabet $\set{0,1}$ and input words are interpreted as the characteristic  sequence of picked numbers minus $T$, i.e.,
the weighted accumulated over a word $w \in \set{0,1}$ equals the sum of $a_i$ such that $i \in \{1, \ldots, k\}$ and $w[i] = 1$ with $T$ subtracted.
One can easily construct such an automaton with $k+2$ states $q_0, \ldots, q_{k+1}$: it moves from $q_i$ to $q_{i+1}$ regardless of the letter if $i \leq k-1$; the transition over $1$ have weight $a_{i+1}$ and 
the transition over $0$ has weight $0$. Then, from $q_k$ it moves to $q_{k+1}$ with both transitions of the weight $-T$.
Finally, in $q_{k+1}$ it has self-loops of the weight $0$.

Next, the automaton $\aut_2$ has the same structure as $\aut_1$, but the last weight is $-T+1$ rather than $-T$.
Observe that if there is a word $w$ distinguishing $\lang((\aut_1, S))$ and $\lang((\aut_2, S))$, then it has to have the value $0$ in $\aut_1$ and $1$ in $\aut_2$ --- since the values of the two automata differ by $1$ and the values of $\aut_1$ are even.
So the two automata are not observationally equivalent exactly when 
the word $w$ encodes the solution for the considered instance of the subset sum problem. \qed
\end{proof}

The subset sum problem has a pseudo-polynomial time algorithm and hence 
the hardness result from Lemma~\ref{l:eq-hard} relies on weights having exponential values w.r.t. the automata sizes. 
Assuming unary weights in automata and the interval endpoints leads to a polynomial-time algorithm for equivalence of PODWA.
More precisely, a PODWA $(\aut, S)$ is \emph{unary} if weights in $\aut$ and interval ends in $S$ are represented in unary.

\begin{theorem}\label{th:equivalence-small-weights-polynomial}
The equivalence problem is \coNP-complete for PODWA and in \PTIME{} for unary PODWA.
\end{theorem}
\begin{proof}
\newcommand{\vass}{\mathcal{V}}
The lower bound for the binary case follows from Lemma~\ref{l:eq-hard}. 
For the upper bound, 
we show that PODWA equivalence reduces to $\Z$-reachability in 2-dimensional vector addition systems (VASS), i.e., 
reachability in which values of counters may become negative.
The weights in the resulting VASS are from the weighted automata. 
The $\Z$-reachability problem for fixed-dimension VASS is \NP{}-complete if vectors' values are represented in binary, and it is in \PTIME{} 
if they are represented in unary~\cite{DBLP:conf/lics/BlondinFGHM15}.

First, consider PODWA  $\podwa_1 = (\aut_1, S_1)$ and $\podwa_2 = (\aut_2, S_2)$. 
If they are not equivalent, then there is $i \neq j$ and a word $w$ such that $\aut_1(w)$ belongs to an $i$-th interval and $\aut_2(w)$ belongs to a $j$-th interval. 
Without loss of generality, $i < j$ and hence  there are values $\lambda_1, \lambda_2$ such that $\aut_1(w) < \lambda_1$ and $\aut_2(w) \geq \lambda_2$. 
There are $|S_1|\cdot |S_2|$ candidates for pairs $\lambda_1, \lambda_2$ and one can verify all pairs. 
Therefore, we assume that $\lambda_1, \lambda_2$ are given and focus on finding $w$ such that $\aut_1(w) < \lambda_1$ and $\aut_2(w) \geq \lambda_2$. 

We construct a VASS $\vass$ of dimension $2$ such that there is a path from the initial state $s_0$ with counters $(0,0)$
to the final state $t$ with counters $(0,0)$ if and only if there is a word $w$ such that $\aut_1(w) < \lambda_1$ and $\aut_2(w) \geq \lambda_2$. 
The VASS $\vass$ is as a product of automata $\aut_1$ and $\aut_2$, where each transition is labeled by a vector of the weights of the corresponding transitions in $\aut_1$ and $\aut_2$.
The $\vass$ has an additional sink state $t$, which is the terminal state, such that from any other state one can reach $t$ over a transition labeled by $(-\lambda_1+1, -\lambda_2)$. 
Additionally, $t$ has self-loops labeled by $(1,0)$ and $(0,-1)$.
Finally, the initial state $s$ of $\vass$ is the pair consisting of initial states of $\aut_1$ and $\aut_2$.

Formally, for $i=1,2$ let $\aut_i = \tuple{\Sigma, Q_i, q_{0,i}, \delta_i, \cost_i}$. The VASS $\vass = \tuple{Q,q_0,\tau}$ is defined as follows:
$Q = Q_1 \times Q_2 \cup \set{t}$, 
$q_0 = \tuple{q_{0,1}, q_{0,2}}$, and
$\tau \subseteq Q \times \Z^2 \times Q$ consist of three types of tuples:
\begin{itemize}
\item tuples $\tuple{(q,s), x, (q',s')}$, for all $q,q' \in Q_1, s,s' \in Q_2$ 
such that 
there exists $a \in \Sigma$ satisfying 
$\delta_1(q,a) = q'$, $\delta_1(s,a) = s'$, and $x = \tuple{\cost_1(q,a,q'), \cost_2(s,a,s')}$
\item tuples $\tuple{(q,s), (-\lambda_1+1, -\lambda_2), t}$, for all $q \in Q_1, s \in Q_2$, and 
\item tuples $\tuple{t,(1,0),t}$ and $\tuple{t,(0,-1),t}$.
\end{itemize}

Now, assume that there is a word $w$ such that $\aut_1(w) < \lambda_1$ and $\aut_2(w) \geq \lambda_2$. 
Then we construct a path in $\vass$ corresponding to $w$, which leads from $s$ with counter values $(0,0)$ to some state with counter values $(a,b)$, where $a < \lambda_1$ and $b \geq \lambda_2$.
Since weights are integers, $a \leq \lambda_1 - 1 $. 
Next, we take a transition to $t$ and the counter values change to $(a',b')$ such that $a' \leq 0$ and $b' \geq 0$. Finally, we can reach counter values $(0,0)$ by taking self-loops over $t$ labeled by $(1,0)$ and $(0,-1)$.
Conversely, consider a path $\pi$ in $\vass$ from $s$ with counter values $(0,0)$ to $t$ with counter values $(0,0)$. Then, let $s'$ be the last state before reaching $t$ and $(x,y)$ be the counter values at that position.
Observe that $x \leq \lambda_1 -1$ and $y \geq \lambda_2$ and hence the prefix of $\pi$ up to $s'$ with $(x,y)$ corresponds to a word $w$ such that $\aut_1(w) < \lambda_1$ and $\aut_2(w) \geq \lambda_2$.
\qed
\end{proof}

\subsection{Unary weights}
Theorem~\ref{th:equivalence-small-weights-polynomial} suggests that restricting the attention to unary PODWA can make learning feasible. 
However, below we show that minimization of automata with bounded weights from $\set{-1, 0, 1}$ may involve exponential-blow up weights, i.e., 
the decrease in the number of states is possible only through introduction of weights of exponential value:
\begin{figure}[!t]
\centering
\begin{tikzpicture}[->,anchor=center]
\node (a) at (-2.8, 0) {a)};
\node[state, initial left] (q01) {$q_0$};

\node[state, below of=q01, yshift=-1cm] (q11) {$q_{1}$};
\node[state, left of=q11, xshift=-1.2cm] (q10) {$q_{1}^{a}$};
\node[state, right of=q11, xshift=1.2cm] (q12) {$q_{1}^{b}$};

\node[state, below of=q11, yshift=-1cm] (q21) {$q_2$};
\node[state, below of=q10, yshift=-1cm] (q20) {$q_{2}^{a}$};
\node[state, below of=q12, yshift=-1cm] (q22) {$q_{2}^{b}$};

\node[state, below of=q21, yshift=-1cm] (q31) {$q_3$};
\node[state, below of=q20, yshift=-1cm] (q30) {$q_{3}^{a}$};
\node[state, below of=q22, yshift=-1cm] (q32) {$q_{3}^{b}$};

\node[below of=q30, yshift=-1cm,inner sep=0.4cm] (d0) {\dots};
\node[below of=q31, yshift=-1cm,inner sep=0.4cm] (d1) {\dots};
\node[below of=q32, yshift=-1cm,inner sep=0.4cm] (d2) {\dots};

\node[below of=q30, yshift=-1.9cm] (fd0) {\dots};
\node[below of=q31, yshift=-1.9cm] (fd1) {\dots};
\node[below of=q32, yshift=-1.9cm] (fd2) {\dots};

\node[state, below of=d1, yshift=-1cm] (q41) {$q_{n}$};
\node[state, below of=d0, yshift=-1cm] (q40) {$q_{n}^{a}$};
\node[state, below of=d2, yshift=-1cm] (q42) {$q_{n}^{b}$};

\node[state, below of=q41, yshift=-1cm] (qs) {$s$};

\draw (qs) edge[loop below] node[right]{$*:0$} (qs);

\draw (q01) edge node[left]{$i:0$} (q11);
\draw (q11) edge node[left]{$i:0$} (q21);
\draw (q21) edge node[left]{$i:0$} (q31);
\draw (q31) edge node[left]{$i:0$} (d1);

\draw (q41) edge node[left,align=center]{$i:0$\\$a:1$\\$b:-1$} (qs);

\draw (q01) edge node[above,sloped]{$a:1$} (q10);
\draw (q11) edge node[above,sloped]{$a:1$} (q20);
\draw (q21) edge node[above,sloped]{$a:1$} (q30);
\draw (q31) edge node[above,sloped]{$a:1$} (d0);

\draw (q12) edge node[right]{$a:0$} (q22);
\draw (q22) edge node[right]{$a:0$} (q32);
\draw (q32) edge node[right]{$a:0$} (d2);
\draw (q42) edge[bend left] node[above,sloped]{$a:1$} (qs);

\draw (q01) edge node[above,sloped]{$b:-1$} (q12);
\draw (q11) edge node[above,sloped]{$b:-1$} (q22);
\draw (q21) edge node[above,sloped]{$b:-1$} (q32);
\draw (q31) edge node[above,sloped]{$b:-1$} (d2);

\draw (q10) edge node[left]{$b:0$} (q20);
\draw (q20) edge node[left]{$b:0$} (q30);
\draw (q30) edge node[left]{$b:0$} (d0);
\draw (q40) edge[bend right] node[above,sloped]{$b:1$} (qs);

\begin{scope}[shift={(6,0)}]
\node (b) at (-2.8, 0) {b)};
\node[state, initial left] (q01) {$q_0$};

\node[state, below of=q01, yshift=-1cm] (q11) {$q_{1}$};

\node[state, below of=q11, yshift=-1cm] (q21) {$q_2$};

\node[state, below of=q21, yshift=-1cm] (q31) {$q_3$};

\node[below of=q31, yshift=-1cm, inner sep=0.4cm] (d1) {\dots};

\node[state, below of=d1, yshift=-1cm] (q41) {$q_{n}$};

\node[state, below of=q41, yshift=-1cm] (qs) {$s$};

\draw (qs) edge[loop below] node[right]{$*:0$} (qs);

\draw (q01) edge node[right,align=left]{$a:2^n$\\ $b:-2^n$\\$i:0$} (q11);
\draw (q11) edge node[right,align=left]{$a:2^{n-1}$\\ $b:-2^{n-1}$\\$i:0$} (q21);
\draw (q21) edge node[right,align=left]{$a:2^{n-2}$\\ $b:-2^{n-2}$\\$i:0$} (q31);
\draw (q31) edge node[right,align=left]{$a:2^{n-3}$\\ $b:-2^{n-3}$\\$i:0$} (d1);
\draw (d1) edge node[right,align=left]{$a:2^{1}$\\ $b:-2^{1}$\\$i:0$} (q41);
\draw (q41) edge node[right,align=left]{$a:2$\\ $b:-2$\\$i:0$} (qs);
\end{scope}
\end{tikzpicture}
\caption{a) The automaton $\podwa_n$. The omitted edges lead to $s$ with weight $0$. b) A minimal automaton equivalent to $\podwa_n$.}\label{f:minautn}\label{f:autn}
\end{figure}
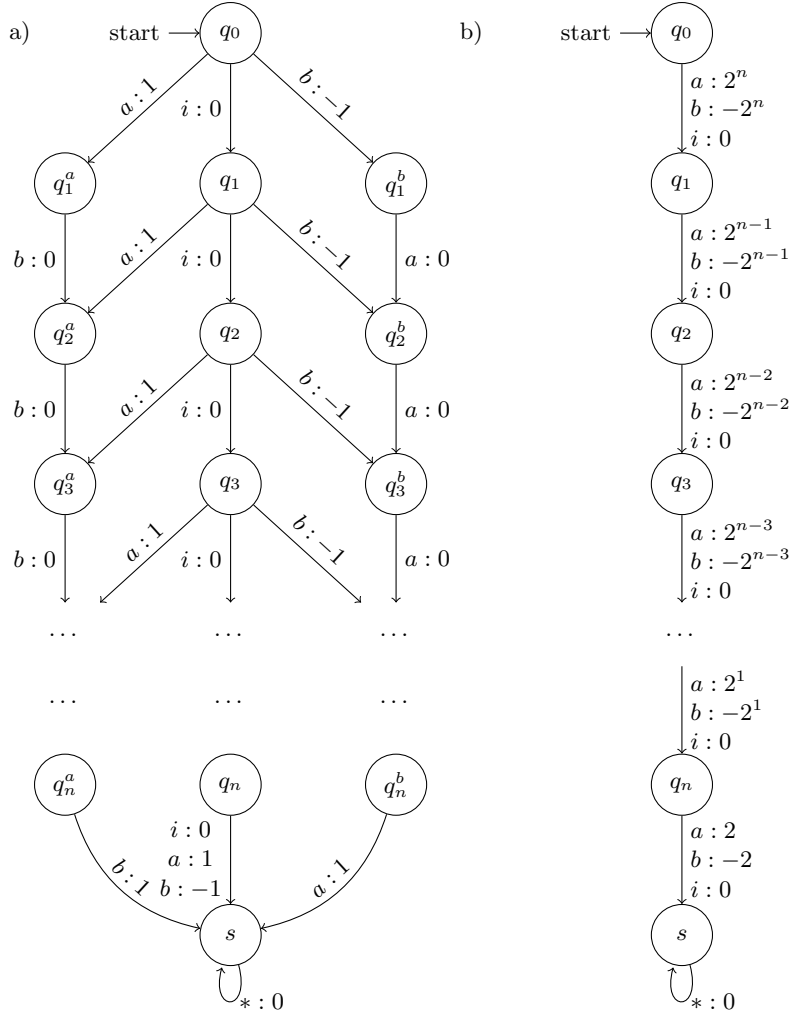

\begin{restatable}{theorem}{MinimalPodwa}
There exists a sequence of PODWA $\podwa_n = (\aut_n, S)$, for $n>1$, with weights $-1, 0, 1$ such that for all $n>1$
every PODWA $(\autB,S)$ equivalent to $\podwa_n$ with $\autB$ having the minimal number of states, has exponential weights in $n$.\label{t:MinimalPodwa}
\end{restatable}
\begin{proof}
\newcommand{\relRC}{\equiv^{0}}
We define, for each $n>1$, a PODWA $\podwa_n = (\aut_n, \set{(-\infty, 0), [0, 0], (0, +\infty)})$ over 
$\Sigma = \set{a, b, i}$ with weights $\set{-1,0, 1}$ such that the minimal equivalent PODWA to $\podwa_n$ needs weights exponential in $n$.

The automaton $\aut_n$ is depicted in Figure~\ref{f:autn} a). 
Intuitively, the value of the word depends on its first $n+1$ letters. If the word starts with the prefix $i^k a$, where $0 \leq k < n$, then it has the value $+1$ unless it is followed by $b^{n-k}$, in which case its value is $0$ (and symmetrically with $i^kb$ and $-1$). Words $i^k$ have value $0$.

An example of a minimal automaton equivalent to $\podwa_n$ is depicted in Figure~\ref{f:minautn} b). 
To show its minimality, observe that  for $j, k \in \set{0, \dots, n+1}$ s.t. $j<k$, the words $i^j$ and $i^k$ have to lead to different states, because $\lang(\podwa_n)(i^ji^{n-j}a) = 2$ and $\lang(\podwa_n)(i^ki^{n-j}a) = 0$.

There are infinitely many minimal automata equivalent to $\lambda_n$ though. For example, one can multiply all the weights of the automaton in Figure~\ref{f:minautn} b) by 2. We can show that all automata equivalent to $\podwa_n$ with the minimal number of states are structurally isomorphic to the automaton  in Figure~\ref{f:minautn} b); this proof is relegated to the appendix.

In all such automata for any $j<n$ we have $\cost(q_j, a) = - \sum_{k=j+1}^{n} \cost(q_k, b)$ and similarly $\cost(q_j, b) = - \sum_{k=j+1}^{n} \cost(q_k, a)$.
Therefore, one can inductively show that for $j<n-1$ we have $\cost(q_j, a) = - \cost(q_j, b) = 2^{n-j-2} (\cost(q_{n-1}, a)+\cost(q_{n}, a))$. 
Since $\cost(q_{n-1}, a)$ and $\cost(q_{n}, a)$ are both positive (because $i^{n-1}$, $i^n$ have the value $0$ and  $i^{n-1}a$, $i^n a$ have positive values), 
we conclude that the value of $\cost(q_0, a)$ is exponential in $n$.
\qed
\end{proof} 

\subsection{Minimization} 
The $L^*$-algorithm relies on the right congruence relation, which has its natural counterpart for DWA. 
The right congruence relation defines the structure of the minimal DWA (which is unique) and hence the active learning algorithm can be applied to minimize DWA. Observe that minimal-size PODWA need not be unique.

\begin{example}
Consider the two binary PODWA presented in Figure~\ref{f:nonisom}.
They both define the language such that all word have positive values exept for the word $a$, which has a negative value.
Both PODWA are equivalent and minimal; if there was an equivalent PODWA with the underlying DWA of a single state $q$, 
then either $\cost(q, a) \geq 1$, which would contradict the value for $a$, or $\cost(q, a) \leq 0$, which would contradict the value for $aa$.
Clearly, the automata are non-isomorphic.

\begin{figure}[t]
\begin{tikzpicture}[->, anchor=center]
\node[state, initial] (q01) {$q_0$};

\node[state, right of=q01, xshift=1cm] (q11) {$q_{1}$};

\node[state, initial, right of=q11, xshift=3cm] (q02) {$q_{0}$};

\node[state, right of=q02, xshift=1cm] (q12) {$q_1$};

\draw (q01) edge node[above,align=center]{$a:-1$\\$b:2$} (q11);

\draw (q11) edge[loop above] node[above,align=center]{$*: 2$} (q11);

\draw (q02) edge node[above,align=center]{$a:-1$} (q12);

\draw (q02) edge[loop above] node[above,align=center]{$b: 2$} (q02);

\draw (q12) edge[loop above] node[above,align=center]{$*: 2$} (q12);
\node (text) at (4, 2) {$S=\set{(-\infty, 0], (0, +\infty)}$};
\end{tikzpicture}
\caption{Two binary PODWA that are equivalent and minimal but not isomorphic.}\label{f:nonisom}
\end{figure}
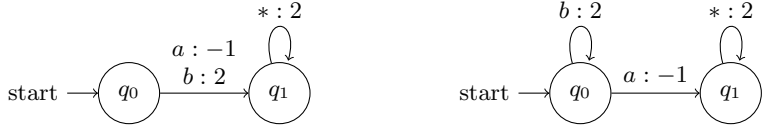
\label{ex:non-equivalent}
\end{example}

\begin{remark}[The right congruence for DWA]
\newcommand{\eqF}{\equiv_f}
For a function $f : \Sigma^* \setminus \{\epsilon\} \to \Z$, consider a relation $\eqF$ defined on non-empty words $w,v$ as follows:
\[ w \eqF v \textrm{ if and only if  for all u $\in \Sigma^*$ we have } f(wu) - f(w) = f(vu) - f(v). \]
The relation $\eqF$ is a counterpart of the right congruence relation for DWA and one can easily show the counterpart of the Myhill-Nerode theorem: 
$f$ is defined by some DWA if and only if $\eqF$ has finitely many equivalence classes, and the relation $\eqF$ defines the structure of the minimal DWA. 
This relation cannot be straightforwardly adapted to PODWA as the result $f(wu) - f(w)$ cannot be inferred from observations for $wu$ and $w$. 
More generally, Example~\ref{ex:non-equivalent} implies that there is no counterpart of $\eqF$ for PODWA as it would imply the uniqueness of 
the structure of minimal PODWA.
\end{remark}

We discuss the complexity of minimization for PODWA, assuming that the set of intervals $S$ is fixed and weights are given in unary. 
We say that DWA $\aut_2$ is \emph{observationally equivalent} to a PODWA $(\aut_1, S)$, if PODWA $(\aut_1, S)$ and $(\aut_2, S)$ are equivalent. 
The $O$-minimization problem is to find a minimal-size DWA $\aut_2$ that is observationally equivalent to a given PODWA $(\aut_1, S)$. 
We study the decision variant of the $O$-minimization problem obtained by stating bound $k$ on $\aut_2$, i.e.,  
 given a PODWA $\podwa= (\aut_1, S)$ and $k>0$, is there a DWA $\aut_2$ with at most $k$ states, which
is observationally equivalent to $\podwa$.

\paragraph{Minimization by merging.} 
A natural approach to minimization of automata is to define an equivalence relation on the set of states of the input automaton $\aut$, 
corresponding to states being \emph{semantically indistinguishable}, and construct the output automaton $\autB$ based on the equivalence classes.
In that approach, semantically indistinguishable are merged into a single state.
Minimization by merging alleviates the problems arising from ambiguity of minimal automata; it guarantees that the input automaton and the minimized one are structurally related. 
We study minimization by merging for PODWA.

A DWA $\autB$ is obtained from a DWA $\aut$ by \emph{merging} if there is 
a surjective (partial) function $f \colon Q_{\aut} \to Q_{\autB}$ from the set of reachable states of $\aut$ onto the set of states $\autB$ such that 
$\delta_{\aut}(q,a) = q'$ if and only if $\delta_{\autB}(f(q),a) = f(q')$. 

The unary $O$-minimization by merging problem is, given an unary PODWA $(\aut, S)$ and $k>0$, is there a DWA $\autB$, with at most $k$ states and (the absolute value of) weights bounded by the weights of $\aut$,  obtained by merging from $\aut$ that is 
observationally equivalent  to $(\aut, S)$.

\begin{theorem}
The unary $O$-minimization by merging problem is $\NP$-complete.
\end{theorem}
\begin{proof}
\newcommand{\autG}{\aut_G}
The problem is in $\NP$ as one can non-deterministically pick a weighted automaton with unary weights $\aut'$ along with the homomorphism witnessing that $\aut'$ can be obtained by merging from $\aut$.
Next, we can check observational equivalence of $\aut$ and $\aut'$ in polynomial time (Theorem~\ref{th:equivalence-small-weights-polynomial}).

We show $\NP$-hardness via reduction from the $k$-coloring problem. Let $G = (V,E)$ be a graph -- for readability we assume it is a directed graph. We construct a binary PODWA $\podwa_G = (\autG, \set{(-\infty, 0], (0, +\infty)})$, which can be $O$-minimized to an automaton with $k+2$ states if and only if 
the vertices of $G$ can be colored with $k$ colors such that each edge connects vertices with different colors.

Let $\Sigma = \{ e^+,e^- \mid e \in E \}$ where $E = \set{e_1, \ldots, e_m}$. 
The states of $\autG$ are $q_0, q_f$ and $\set{ q_v : v \in V}$.
For an edge $e_i=(v,u)$ we define $\delta(q_0, e_i^-)=v$ and $\delta(q_0, e_i^+)=u$, i.e., over $e_i^-, e_i^+$ the automaton reaches both ends of $e$. All the remaining transitions lead to $q_f$.

We define weights function $\cost$ so that pairs of states $q_v, q_u$ can be merged if and only if they correspond to vertices $u,v$ not connected in $G$.
For any $e \in E$ we will ensure that in $\autG$ the values of words $e^- e^-, e^+ e^+$ are negative and the value of words $e^- e^+, e^+ e^-$ are positive.
This guarantees that $e^+$ and $e^-$ cannot lead to the same state.
Intuitively, after $e^-$ the state in $\autG$ has outgoing transitions over $e^-, e^+$, where the weight of $e^+$ is strictly greater than the weight of $e^-$, and
for the state reachable over $e^+$, the order of weights is the opposite.

For every $e_i = (v, u) \in E$ we define $\cost(q_0,e_i^-) = \cost(q_0,e_i^+)=-3i-1$. Then, for $q_v$ we define $\cost(q_v,e_i^-)=3i$ and $\cost(q_v,e_i^+)=3i+2$. For $q_u$ we define 
$\cost(q_u,e_i^-)=3i+2$ and $\cost(q_u,e_i^+,q_f)=3i$.
For $u$ that is not an endpoint of $e_j$ we set $\cost(q_u, e_j^-)=\cost(q_u, e_j^+) = 3j+1$. The weights $\cost(q_f, *)$ are all $0$.

We show that $G$ is $k$-colorable if and only if $\podwa_G$ can be $O$-minimized to an automaton with $k+2$ states.
First, observe that the values $e_i^- e_i^-$ and $e_i^+ e_i^+$ in $\autG$ are $-1$ and 
the values $e_i^- e_i^+$ and $e_i^+ e_i^-$ are $1$ and hence $q_u$ and $q_v$ cannot be merged. 
Second, $q_0$ and $q_f$ cannot be merged with one another or any other state; all words starting from $q_0$ are negative, and
all word starting from $q_0$ retain their values. No other state has such a property. 
Therefore, if $\autG$ is minimized by merging to an automaton with $k+2$ states, then $k$ is at least equal to the chromatic number of $G$.

Conversely, assume that $\lambda \colon V \to \set{1, \ldots, k}$ is a valid coloring of $G$.
We construct a DWA $\autG'$ with the same structure as $\autG$, with the property that states corresponding to nodes of the same color have the same values of outgoing edges.
Recall that 
for $u$ that is not an endpoint of $e_j$ we set $\cost(q_u, e_j^-)=\cost(q_u, e_j^+) = 3j+1$.
Changing any such weight to $3j$ or $3j+2$ leads to an equivalent automaton. 
Indeed, the state $q_u$ can be reached with values $-3i-1$, where $i \neq j$ and hence
the values $-3i-1 + 3j, -3i-1 + 3j +1, -3i-1 + 3j +2$ are either all positive or all negative. 
With that observation, we can modify weights in $\autG$ such that for $u, v$ with the same color, the weights of all outgoing transitions from $q_i, q_v$ are the same and hence the states can be merged. 
Assume that $u[1], \ldots, u[k]$ have the same color; then for every edge $e$ at most one of these vertexes can be an endpoint of $e$; if there is such $u[i]$ then
we fix weights of all transitions $(q_{u[1]}, e^-), \ldots, (q_{u[k]}, e^-)$ to be the same as the weight of $(q_{u[i]}, e^-)$. 
If there is no such vertex, we do not change the weights. We fix weights over $e^+$ accordingly.
Observe, the in the resulting automaton states $q_{u[1]}, \ldots, q_{u[k]}$ have all the outgoing transitions to $q_f$, and 
transitions over the same letter have the same weight. Therefore, they all can be merged into the same state. 
\qed
\end{proof}

\section{Conclusions}
This paper introduces partially-observable deterministic weighted automata, which address the difficulty in specification synthesis originating from the need of
feeding the exact values to the specification procedure.
We have studied the basic properties of the model as well as problems related to specification synthesis via automata learning: equivalence and minimization. 
The main contribution of the paper is identifying obstacles in developing polynomial-time active learning algorithm for the new model.
While our framework is unlikely to admit such an algorithm, it is possible that restricting the equivalence notion may lead framework admitting 
polynomial-time active learning algorithm.

\subsection*{Acknowledgements}
This work was supported by the National Science Centre (NCN),
Poland under grant 2020/39/B/ST6/00521.
 
The Version of Record of this contribution (without the appendix) is published in LNAI,volume 14281, and is available online at the following URL: \url{https://doi.org/10.1007/978-3-031-43619-2\_52} . 
\bibliographystyle{plain}
\bibliography{final}

\appendix
\section{Appendix}



\MinimalPodwa*
\begin{proof}
Here we only fill the remaining details of the proof presented in the main body of the paper.
For readability, we will say ``the value $(-\infty, 0)$ / $[0, 0]$ / $(0, +\infty)$'' rather than the technically correct ``the value $0 / 1 / 2$''.

Assume that $\aut_w$ is an minimal automaton observationally equivalent to $\podwa_n = (\aut_n, S)$. 
We have already argued that $\aut_w$ has $n+2$ states: the initial state $q_0$,  states $q_j$, 
for $j \in \set{1, \dots, n}$, reachable over the words $i^j$ and the state $s$ reachable over $i^{n+1}$.
Here we argue that the remaining transitions of $\aut_w$ are as in Figure~\ref{f:minautn} b).

Observe that from any state $q_j$ for $j\leq n$ there is a transition over $a$ with a positive weight (so that the word $i^ja$ is in $(0, +\infty)$) and a transition over $b$ with a negative weight (so that the word $i^jb$ is in $(-\infty, 0)$).

The transitions from $s$ can only lead to $s$: note that all words starting with $i^{n+1}$ have the value $[0, 0]$ in $\podwa_n$. 
If the word $i^{n+2}$ led to a state $q_j$ for $j n+1$, then $i^{n+2}a$ would have the value $(0, +\infty)$ by the above observation.
It also follows that the weight of all edges from $s$ (to itself) is $0$.

It remains to show that the for any $j \leq n$ we have $\delta(q_j, a) = q_{j+1}$ and  $\delta(q_j, b) = q_{j+1}$. 

First, we show that 
\begin{equation} 
\mathrm{for\ all\ } j<k \leq n \mathrm{\ in\ } \aut_w \mathrm{\ we\ have\ }  \delta(q_j, a) \neq \delta(q_k, a) \label{p:diff}
\end{equation}
Assume towards contradiction that $\delta(q_j, a) = \delta(q_k, a)$. 
Consider $w_1 = i^j a$ and $w_2 = i^k a$ such that $j<k$ and $q= \hat{\delta}(q_0, w_1) = \hat{\delta}(q_0, w_2)$. 
We show that $\hat{\delta}(q, b^{n-k}) = s$. 
This is because the value of $w_1 b^{n-k}$ and $w_2 b^{n-k} a$ in $\podwa_n$ is both $[0, 0]$ and 
$s$ is the only state where the weight of the transition $a$ is $0$.

On the other hand, the value of $\aut_w$ for $w_1 b^{n-k}$ is in $(0, +\infty)$. 
Since $\hat{\delta}(w_1 b^{n-k})$ is $s$, this means that the value of $\aut_w$ for $w_1 b^{n-k} b^{k-j}$ is in $(0, +\infty)$. 
But the value of $\aut_b$ for this word is in $[0,0]$, which contradicts the equivalence.

We now show that
\begin{equation} 
\mathrm{for\ all\ } j \mathrm{\ we\ have\ }\delta(q_j, a) \neq q_0 \mathrm{\ and\ } \delta(q_j, b) \neq q_0  \label{p:start}
\end{equation}

Assume w.l.o.g. that $\delta(q_j, a)=q_0$. 
Observe that $\hat{\delta}(q_0, i^jab^{n-j}) = s$. Let $Y$
be the set of states along the run over $i^jab^{n-j}$, i.e., $Y = \set{\hat{\delta}(q_0,w) \mid \exists v \in \Sigma^*. wv = i^jab^{n-j}}$ and 
\[
 X = \set{q_0, \dots, q_{n},s} \setminus Y
\]

Observe that  $\hat\delta(q_0, i^ja)=q_0$, i.e., the state $q_0$ occurs at least twice  in  the run over  $i^jab^{n-j}$ and hence
the set $Y $ has at most $n+1$ states. 
Therefore,  $X$ is non-empty and does not contain $s$ as $s \in Y$. 
Let $q_G$ be the state with the greatest index in $X$. 
Since some state from $Y$ has a transition over $b$ to $s$, (\ref{p:diff}) implies that for $q_G \notin Y$ we have
$\delta(q_G, b) \neq s$. Thus, $\delta(q_G, b) =q_L$ for some $L\leq G$. 

Now, consider the words of the form $w_c = i^j a i^G (b i^{G-L})^c$ for $c>0$.
The value of $\podwa_n$ for each $w_c$ is either $(0, +\infty)$ or $[0, 0]$. 
However, since all the weights for $i$ are $0$, and the transition from $q_G$ over $b$ has a negative weight, 
for some large enough $c$ the value of $\aut_w$ for $w_c$ will be negative -- a contradiction.

We finally show that for each for each $k \leq n$ we have $\delta(q_k, a) = \delta(q_k, b) = q_{k+1}$ and 
$\delta(q_n, a) = \delta(q_n, b) = s$.
From (\ref{p:diff}) and (\ref{p:start}) it follows that the states of $\aut_w$ after reading words $a$, $ia$, \dots, $i^na$ are some permutation of the states $q_1, \dots, q_{n}, s$ -- and the same for $b$. 

Assume w.l.o.g. that for some $k < n$ we have $\delta(q_k, a) \neq q_{k+1}$ (resp., $\delta(q_n, a) \neq s$). 
It means that there is $l < n$ such that $\delta(q_l, a) = q_m$ for $m \leq l$. 
Also, there is a state $q_s$ such that $\delta(q_s, b)= q_m$.

Now consider the word
\[ w_c = i^s b (a i^{m-l})^c\]

Notice that for each $c$, we have that $\hat\delta(w_c) = q_l$ and $c(q_l, a) > 0$. It follows that for sufficiently large $c$, the value of $\aut_w$ for $w_c$ is in $(0, +\infty)$ -- which is a contradiction with the value of $\aut_n$.

\end{proof}

\end{document}